\documentclass[hyper]{JHEP3}

\usepackage[dvips]{graphicx}
\usepackage{epsfig,multicol,amssymb,amsmath,cancel}

\newcommand\fverb{\setbox\fverbbox=\hbox\bgroup\verb}
\newcommand\fverbdo{\egroup\medskip\noindent%
            \fbox{\unhbox\fverbbox}\ }
\newcommand\fverbit{\egroup\item[\fbox{\unhbox\fverbbox}]}
\newbox\fverbbox

\def\bea{\begin{eqnarray}}
\def\eea{\end{eqnarray}}
\def\beq{\begin{equation}}
\def\eeq{\end{equation}}

\title{Mixed Mediation of Supersymmetry Breaking with Anomalous
$U(1)$ Gauge Symmetry}

\author{Kiwoon Choi,$^a$ Kwang Sik Jeong,$^b$ Ken-Ichi Okumura$^c$
and Masahiro Yamaguchi$^b$ \\
$^a$Department of Physics, KAIST, Daejeon 305-701, Korea \\
$^b$Department of Physics, Tohoku University, Sendai 980-8578, Japan  \\
$^c$Department of Physics, Kyushu University, Fukuoka 812-8581, Japan \\
E-mail : \email{kchoi@kaist.ac.kr},
\email{ksjeong@tuhep.phys.tohoku.ac.jp},
\email{okumura@phys.kyushu-u.ac.jp},
\email{yama@tuhep.phys.tohoku.ac.jp} }

\preprint{KYUSHU-HET 125, TU-882}

\abstract{
Models with anomalous $U(1)$ gauge symmetry contain
various superfields which can have nonzero supersymmetry breaking
auxiliary components providing the origin of soft terms in the
visible sector, e.g. the $U(1)$ vector superfield, the modulus or
dilaton superfield implementing the Green-Schwarz anomaly
cancellation mechanism, $U(1)$-charged but standard model singlet
matter superfield required to cancel the Fayet-Iliopoulos term, and
finally the supergravity multiplet. We examine the relative strength
between these supersymmetry breaking components in a simple class of
models, and find that various different mixed mediations of supersymmetry
breaking, involving the modulus, gauge, anomaly and $D$-term
mediations, can be realized depending upon the characteristics of $D$-flat
directions and how those $D$-flat directions are stabilized
with a vanishing cosmological constant.
We identify two parameters which represent such properties and thus
characterize how the various mediations are mixed.
We also discuss the moduli stabilization and soft terms in a variant
of KKLT scenario, in which the visible sector K\"ahler modulus is
stabilized by the $D$-term potential of anomalous $U(1)$ gauge symmetry.
}


\begin{document}

\section{\label{s1}Introduction}

Anomalous $U(1)$ gauge symmetry, which will be referred to as $U(1)_A$
in the following, appears  often in 4-dimensional (4D) effective
theory of string compactification.
It accompanies a modulus $T$ which transforms nonlinearly under $U(1)_A$,
and the holomorphic gauge kinetic function of the model depends on $T$
as $f_a\ni k_aT$, where $k_a$ is a real constant.
Then the variation of $f_a$ under $U(1)_A$ cancels the anomaly due to
the loops of light fermions, realizing the Green-Schwarz (GS) anomaly
cancellation mechanism \cite{GS-mech}.
The $U(1)_A$ gauge boson receives a mass through the St\"uckelberg
mechanism associated with the nonlinear transformation of $T$,
which is typically not far below the string or Planck scale.
The nonlinear transformation of $T$ also induces a moduli-dependent
Fayet-Iliopoulos (FI) term \cite{FI-string}.
It has been noticed that anomalous $U(1)$ gauge symmetry can have
a variety of interesting phenomenological implications.
It can be used to forbid dangerous interactions such as the ones which
lead to a too rapid proton decay, or to explain the smallness of
some couplings in the low energy theory.
In some cases, it can be identified as a flavor symmetry that explains
the observed hierarchical fermion masses
\cite{FN-mech,U1A-fermion-mass,u1a-yukawa1,u1a-yukawa2}.

Due to the existence of the GS modulus $T$ and the associated FI term,
anomalous $U(1)$ gauge symmetry can play an important role in supersymmetry
(SUSY) breaking and its transmission to the supersymmetric standard model
\cite{U1A-susy-breaking,u1a-sb1,u1a-sb2,D-U1A,GS-Stabilization,u1a-soft-terms,
Gravity-gauge-U1A,GS-Kahler,KKLT-U1A,U1A-LVS}.
In most cases, the FI term has a vacuum expectation value (VEV)
far above the weak scale, even close to the Planck scale in some cases.
Then, to avoid a too large $D$-term SUSY breaking, the FI term should be
cancelled by other contribution to the $U(1)_A$ $D$-term due to $U(1)_A$
charged but standard model (SM) singlet matter field $X$.
This SM singlet $X$ can play another important role.
In many cases, string models with $U(1)_A$ contain exotic matter fields
$\Phi$, $\Phi^c$ which are vector-like under the SM gauge group,
and these exotic matter fields get a mass far above the weak scale through
the Yukawa coupling to $X$.
Then they can be identified as the messenger for gauge-mediated SUSY breaking
if the $F$-component of $X$ develops a nonzero VEV.

Therefore, models with anomalous $U(1)$ gauge symmetry contain
various sources of SUSY breaking, including {\it (i)} the $U(1)_A$
D-term, {\it (ii)} the $F$-component of the GS modulus $T$, {\it
(iii)} the $F$-component of the chiral matter superfield $X$ whose
lowest component cancels the FI term and provides a large mass to
exotic matter fields, and finally {\it (iv)} the auxiliary component
of the supergravity (SUGRA) multiplet which is generically of the
order of the gravitino mass $m_{3/2}$. Then the visible sector soft
terms receive modulus-mediated contribution of the order of $F^T$
and gauge-mediated contribution of the order of
$\frac{g^2}{8\pi^2}\frac{F^X}{X}$, as well as the anomaly-mediated
contribution of the order of $\frac{g^2}{8\pi^2}m_{3/2}$.
Furthermore, there can be $D$-term contribution to scalar masses for
$U(1)_A$ charged matter fields. This means that the four well-known
mediation schemes of SUSY breaking, i.e. moduli mediation
\cite{Moduli-mediation}, gauge mediation
\cite{gauge-mediation-classic,Gauge-mediation}, anomaly mediation
\cite{Anomaly-mediation} and $D$-term mediation, generically appear
together in models with $U(1)_A$.

In this paper we wish to examine the possible pattern of the mediation
of SUSY breaking in models with anomalous $U(1)$ gauge symmetry.
As we will see, the relative strength between different mediations
crucially depends on the characteristics of the $D$-flat directions,
and also on how the $D$-flat directions are stabilized.
Depending upon the detailed form of the K\"ahler potential and
superpotential, various different mixed mediations involving some or
all of the moduli, gauge, anomaly and $D$-term mediations can be
realized within a relatively simple class of models.

This paper is organized as follows.
In section 2, we discuss generic features of supersymmetry breaking and
the resulting pattern of soft terms in models with anomalous $U(1)$
gauge symmetry.
In section 3, we consider a set of specific models to examine
the stabilization of $D$-flat directions under the constraint of nearly
vanishing cosmological constant, and evaluate the relative strength between
different mediations in each model.
Section 4 is the conclusion.

\section{\label{s2}SUSY breaking in models with anomalous $U(1)$}

In this section, we discuss generic aspects of supersymmetry breaking
in models with anomalous $U(1)$ symmetry.
We first examine the relations between different SUSY breaking auxiliary
components in models with $U(1)_A$, and then discuss the resulting
pattern of soft terms.

\subsection{\label{ss2.1}SUSY breaking auxiliary components}

In the presence of anomalous $U(1)$ gauge symmetry, the quantum
consistency of the theory is ensured by the GS anomaly cancellation
mechanism \cite{GS-mech}.
This mechanism is implemented by a non-linear variation of the GS modulus
\bea
T \rightarrow T - \frac{\delta_{GS}}{2}\Lambda_A
\eea
under the gauge transformation
\bea
V_A \rightarrow V_A -\frac{1}{2}(\Lambda_A+\Lambda^*_A),
\eea
where $V_A$ is the vector superfield containing the $U(1)_A$ gauge field.
For the anomaly cancellation to work, the holomorphic gauge kinetic
function of the model should contain a $T$-dependent piece,
\bea
f_a= k_a T + \cdots,
\eea
where $k_a$ is a real constant and the ellipsis stands for
the $T$-independent part.
In the normalization convention of $T$ for which $k_a={\cal O}(1)$,
the anomaly cancellation implies
\bea
\frac{\delta_{GS}}{2} = {\cal O}\left(\frac{1}{8\pi^2}\right).
\eea
Since the $U(1)_A$ invariance forces the modulus K\"ahler potential $K_0$
to be a function of the gauge-invariant combination
$t_A = T+T^*-\delta_{GS}V_A$, the GS mechanism dynamically induces
a modulus-dependent FI term\footnote{
Unless specified, we will use the convention $M_{Pl}=1$ throughout
this paper, where $M_{Pl}=2.4\times 10^{18}$ GeV is the reduced
Planck scale.
}
\bea
\label{fi}
\xi_{FI} = \frac{\delta_{GS}}{2} \partial_TK_0,
\eea
while rendering the vector superfield $V_A$ massive through
the St\"uckelberg mechanism:
\bea
\label{stuk}
\Delta M_V^2 =\frac{g_A^2\delta^2_{GS}}{2} \partial_T\partial_{\bar T}K_0,
\eea
where $g_A$ is the $U(1)_A$ gauge coupling.

To proceed, let us consider 4D effective SUGRA model with chiral
superfields $\Phi_I=\{T_M,\phi_i\}$, where $T_M=\{T,T_\alpha\}$
stand for generic moduli including the GS modulus $T$ and $\phi_i$
are chiral matter superfields with $U(1)_A$ charge $q_i$.
Under $U(1)_A$, these chiral superfields transform as
\bea
\delta\Phi_I =\eta^I\Lambda_A,
\eea
where the holomorphic Killing vectors $\eta^I$ are given by
\bea
\eta^T = -\frac{1}{2}\delta_{GS}, \quad \eta^{T_\alpha}=0, \quad
\eta^{\phi_i}=q_i\phi_i.
\nonumber
\eea
Since the moduli stabilization is relatively insensitive to the form
of the matter K\"ahler metric, we assume for simplicity that the
K\"ahler metric of $\phi_i$ is independent of moduli.
(In fact, most of our results apply well at least qualitatively
to the case when the matter K\"ahler metrics are moduli-dependent.)
Then the K\"ahler potential of the model takes the form:
\bea
K = K_0(t-\delta_{GS}V_A,T_\alpha,{T}^*_\alpha)
+ \sum_i{\phi}^*_i e^{2q_iV_A}\phi_i,
\eea
where $t=T+T^*$ and we have ignored the terms of higher order in $\phi_i$
which are presumed to be suppressed by $1/M_{Pl}$.
The associated $U(1)_A$  gauge boson mass and $D$-term are given by
\bea
\label{dterm}
M_V^2 &=& 2g_A^2\eta^I\eta^J\partial_I\partial_{\bar{J}}K
= 2g_A^2\left( M_{GS}^2+\sum_i q^2_i|\phi_i|^2\right),
\nonumber \\
D_A &=& -\eta^I\partial_I K = \xi_{FI}-\sum_iq_i|\phi_i|^2,
\eea
where
\bea
\xi_{FI} = \frac{\delta_{GS}}{2}\partial_T K_0,\quad
M_{GS}^2=\frac{\delta_{GS}^2}{4}\partial_T
\partial_{\bar T}K_0.
\nonumber
\eea
From this, one easily finds
\bea
\label{relations}
m_{3/2}D_A = \eta^IF^{\bar J}\partial_I\partial_{\bar{J}}K= \sum_i
q_i|\phi_i|^2\left(\frac{F^{i}}{\phi_i}\right)^*
-\frac{1}{2}\delta_{GS}\sum_M F^{\bar M}\partial_T
\partial_{\bar M}K_0,
\eea
where $m_{3/2}=e^{K/2}W$ for the superpotential $W$, and the auxiliary
$F$-component of $\Phi_I=\{T_M,\phi_i\}$ is defined as
\bea
F^I=-e^{K/2}K^{I\bar{J}}(\partial_JW +W\partial_JK)^*.
\nonumber
\eea
Combining (\ref{relations}) with the stationary  condition
$\partial_I (V_F+V_D)=0$, one can derive \cite{KKLT-U1A,F-D-U1A}
\bea
\label{D-F-rel-2}
\left(V_F + 2|m_{3/2}|^2 + \frac{1}{2}M^2_V \right)D_A =
-F^IF^{\bar J}\partial_I\partial_{\bar J}(\eta^L\partial_L K)
+ V_D\eta^I\partial_I g^2_A,
\eea
where the $F$ and $D$ term scalar potentials are given by
\bea
V_F = K_{I\bar J}F^I F^{\bar J} - 3e^K|W|^2, \quad
V_D = \frac{g^2_A}{2}D^2_A.
\eea
This relation can be generalized to the case including an uplifting
potential $V_{\rm lift}$ which might be necessary to achieve
a vacuum with nearly vanishing cosmological constant.
As long as $V_{\rm lift}\ll M_V^2$ (in the unit with $M_{Pl}=1$),
which is always the case in models with low energy SUSY, the effect
of $V_{\rm lift}$ in the generalized version of (\ref{D-F-rel-2})
can be safely ignored, and we can apply (\ref{D-F-rel-2}) to the case
with $V_{\rm lift}$ as well.
We then find
\bea
\label{dterm1}
g_A^2 D_A &=&
-\frac{F^IF^{\bar J}\partial_I\partial_{\bar J}
(\eta^L\partial_L K)}{\eta^I\eta^{\bar J}\partial_I\partial_{\bar{J}}K}
\left(1+{\cal O}\left(\frac{m_{3/2}^2}{M_V^2}\right)\right)
\nonumber \\
&=&
\frac{\frac{1}{2}\delta_{GS}\sum_{MN} F^MF^{\bar N}\partial_T\partial_M
\partial_{\bar N}K_0
-\sum_i q_i|F^i|^2}{\frac{1}{4}\delta_{GS}^2\partial_T
\partial_{\bar T}K_0+\sum_i q_i^2|\phi_i|^2}
\left(1+{\cal O}\left(\frac{m_{3/2}^2}{M_V^2}\right)\right).
\eea

In models with low energy SUSY, $m_{2/3}$ and $\sqrt{D_A}$ have a VEV
in TeV or multi-TeV range (or lower than TeV).
On the other hand, although it depends on the stabilization of
$D$-flat directions, typically there exist some $U(1)_A$ charged
(but SM singlet) matter fields having a VEV far above TeV.
Also, with the St\"uckelberg contribution (\ref{stuk}),
the $U(1)_A$ gauge boson mass $M_V$ is rather close to the Planck
scale or the string scale, thus is much higher than TeV.
Then the relations (\ref{dterm}), (\ref{relations}) and (\ref{dterm1})
give rise to
\bea
\label{rel}
&&
\frac{1}{2}\delta_{GS}\partial_T
K_0 \,\simeq\, \sum_i q_i |\phi_i|^2,
\nonumber \\
&&
\frac{1}{2}\delta_{GS}\sum_M F^M \partial_M\partial_{\bar T} K_0
\,\simeq\, \sum_i q_i|\phi_i|^2
\left(\frac{F^i}{\phi_i}\right),
\nonumber \\
&&
g_A^2 D_A \,\simeq \,
\frac{\frac{1}{2}\delta_{GS}\sum_{MN}
F^MF^{\bar N}\partial_T\partial_M\partial_{\bar N}K_0
- \sum_i q_i|\phi_i|^2\left|\frac{F^i}{\phi_i}\right|^2}{\frac{1}{4}
\delta_{GS}^2\partial_T\partial_{\bar T} K_0+\sum_i q_i^2
|\phi_i|^2},
\eea
which in fact correspond to the lowest component, the $F$-component,
and the $D$-component of the equation of motion for the $U(1)_A$ vector
superfield $V_A$ in the limit that $V_A$ receives a large supersymmetric mass.

Eq. (\ref{rel}) includes relations between the moduli $F$-components
$F^M$, the matter $F$-components $F^i$, and the $U(1)_A$ $D$-term.
To see the implication of those relations more clearly, let $X$
denote the $U(1)_A$ charged (but SM singlet) matter field with the
{\it largest} VEV, and consider the case that the FI term in $D_A$
is cancelled dominantly by $q_X|X|^2$, so
\bea
\label{dflat1}
|X|^2 \,\sim\, |\xi_{FI}|\,\gg\, D_A.
\eea
We further assume that $F^T$ and $F^X/{X}$ are at least comparable to
other moduli and matter $F$-components, respectively.
Then, in the convention with $q_X=-1$, we find
\bea
\label{FD}
\frac{F^X}{X} &\simeq &
\left(\frac{K_0^{\prime\prime}}{K_0^\prime}\right)F^T,
\nonumber \\
g^2_A D_A &\simeq &\left(
\frac{\left(\frac{K_0^{\prime\prime}}{K_0^\prime}\right)^2
-\frac{K_0^{\prime\prime\prime}}{K_0^\prime}}{1-\frac{\delta_{GS}}{2}
\frac{K_0^{\prime\prime}}{K_0^\prime}}\right)|F^T|^2,
\nonumber \\
\frac{F^X}{X} &=& -m_{3/2}^*
\left(1+\frac{\partial_{ X}W}{X^* W}\right)^*,
\eea
where the prime denotes the derivative with respect to $t=T+T^*$,
and the last relation is derived from  $F^X=-e^{K/2}K^{X\bar{I}}(D_I W)^*$.
These relations suggest that the relative ratios between the four
SUSY breaking auxiliary components $F^T, F^X, m_{3/2}$ and $D_A$ are
determined mostly by
\bea
\label{r1r2}
R_1 \equiv -\frac{\delta_{GS}}{2}\frac{K_0^{\prime\prime}}{K_0^\prime},
\quad
R_2\equiv \left(\frac{D_XW}{X^*W}\right)^*=1
+ \left(\frac{\partial_X W}{X^*W}\right)^*,
\eea
where $R_1>0$ in our convention with $q_X=-1$.
More specifically, in the basis where $F^T$ is real, we obtain
\bea
\label{ratio_auxil}
F^T : \frac{F^X}{X} : m^*_{3/2}: g_A\sqrt{D_A}
\,\, \simeq \,\,
\frac{\delta_{GS}}{2R_1}:\, -1 \,:\, \frac{1}{R_2}
\,:\frac{1}{\sqrt{R_1+1}},
\eea
where the first three relations are precise, while the relative size
of $\sqrt{D_A}$ is approximately estimated under the assumption that
$K_0^{\prime\prime\prime}/K^{\prime\prime}_0$ is comparable to
(or smaller than) $K_0^{\prime\prime}/K_0^\prime$, which holds
true in most cases.
In the subsections \ref{ss3.2} and \ref{ss3.3}, we will discuss explicit
examples in which the $D$-flatness is achieved as (\ref{dflat1}),
and therefore the ratios between the SUSY breaking auxiliary components are
given by (\ref{ratio_auxil}).

Let us now discuss the possible ranges of $R_1$ and $R_2$.
The first possibility is that $K_0^{\prime\prime} \sim K_0^\prime$,
which would result in
\bea
R_1={\cal O}\left(\delta_{GS}\right)=
{\cal O}\left(\frac{1}{8\pi^2}\right).
\eea
An example for such case is provided by the modulus K\"ahler potential
\bea
K_0 \simeq -n_0\ln(T+T^*-\delta_{GS}V_A)
\eea
for $T$ stabilized at a value of order unity.

Another even more interesting possibility is that $T$ is stabilized
at near a point with $\xi_{FI}=\frac{1}{2}\delta_{GS}K_0^\prime=0$,
which would give
\bea
R_1 ={\cal O}(1)\quad \mbox{or} \quad \gg 1.
\eea
This would be a plausible scenario if the K\"ahler potential admits a limit
with $\xi_{FI} =0$, or more generally a limit with $\xi_{FI}$ far below
$M_{Pl}^2$, since $\xi_{FI}=\phi_i=0$ is a point of enhanced (approximate)
symmetry\footnote{
Note that $D_A=0$ and the global part of $U(1)_A$ is restored at this point.
}
and satisfies the equation of motion in the limit to ignore SUSY breaking
effects.
Then a (local) minimum of the scalar potential with nonzero but small VEV
of $|X|/M_{Pl}$ can be developed by small SUSY breaking effects,
which results in a tiny VEV of $\xi_{FI}/M_{Pl}^2\simeq -|X|^2/M_{Pl}^2$.
It has been known that many brane models constructed within type IIA or IIB
string theory admit supersymmetric brane configuration with smooth background
spacetime geometry, which gives rise to an anomalous $U(1)$ gauge
symmetry with $\xi_{FI}=0$ \cite{uranga}.
In addition, it has recently been noticed that heterotic string compactification
also can give rise to such a solution with $\xi_{FI}=0$ \cite{Stability wall},
which is at the boundary of the K\"ahler moduli space in which the Hermitian
Yang-Mills equations are satisfied.
There are also examples that $\xi_{FI}=0$ correspond to a singular limit
of collapsing cycle (or orbifold) \cite{GS-blowing-up}.
So an anomalous $U(1)$ symmetry with $\xi_{FI}=0$ somewhere in moduli space
is not unusual, but appears quite often in phenomenologically interesting set
of string compactifications.

The value of $R_2$ is determined mostly by the mechanism to stabilize
the $D$-flat directions involving $X$.
In case that $X$ is the dominant matter field which cancels the FI term,
there is only one relevant $D$-flat direction described by
$Xe^{-2T/\delta_{GS}}$.
However, if there exists additional matter field $Y$ having an opposite
$U(1)_A$ charge and comparable VEV, there will be additional $D$-flat
direction described by a $U(1)_A$ invariant holomorphic monomial of $X$
and $Y$.
If the superpotential of the model is independent of these $X$-dependent
$D$-flat directions, so $\partial_XW=0$, by definition $R_2$ has a value
close to the unity.
In this case, the $D$-flat directions should be stabilized by nontrivial
structure of the K\"ahler potential which might be induced by radiative
corrections.
Another possibility is that the superpotential contains a higher dimensional
term depending on $X$, e.g. $\Delta W\sim X^3Y$ for $q_Y=-3q_X$, and then
$X$ is stabilized by the competition between the supersymmetric potential
$|\partial_X W|^2$ and the SUSY breaking terms  controlled by
$m_{3/2}\sim W/M_{Pl}^2$, e.g. an $A$-term of the form $m_{3/2}X\partial_XW$
or a scalar mass term of the form $-m_{3/2}^2|X|^2$.
Such setup stabilizes the $D$-flat direction  at a (local) minimum satisfying
\bea
|\partial_X W|^2\sim m_{3/2} X\partial_XW \quad
\mbox{or}\quad m_{3/2}^2 |X|^2,
\eea
for which
\bea
R_2=1+ \left(\frac{\partial_XW}{X^*W}\right)^* ={\cal O}(1).
\eea

It is also possible that $|R_2|$ has a value much smaller or much
larger than the unity.
For instance, if the $D$-flat direction $Xe^{-2T/\delta_{GS}}$ is stabilized
by a nonperturbative superpotential $\Delta W\sim X^ne^{-2nT/\delta_{GS}}$
at near the supersymmetric solution with $D_XW=0$, we have
\bea
|R_2|\ll 1.
\eea
In other case that the superpotential provides spontaneous SUSY breaking
in the global SUSY limit, e.g. through the Polonyi term of the form
$\Delta W\sim e^{-2T/\delta_{GS}}X$, it is also possible that
\bea
|R_2|\gg 1.
\eea
In this case, the condition for vanishing cosmological constant provides
an upper limit
\bea
|R_2|\sim \left|\frac{\partial_XW}{X W}\right| \lesssim {\cal O}\left(
\frac{M_{Pl}}{\sqrt{|\xi_{FI}|}}\right),
\eea
where we have used $|X|^2\simeq |\xi_{FI}|$ together with the observation
that $|\partial_XW|^2 \lesssim {\cal O}(|W|^2/M_{Pl}^2)$ in order for the
cosmological constant to be nearly vanishing.

The above discussion implies that even within a relatively simple class
models a variety of different patterns of SUSY breaking can be realized,
depending on the characteristics of $D$-flat directions and how those
$D$-flat directions are stabilized.
In more complicate situation in which there exist multiple number of moduli
and/or of $U(1)_A$-charged matter fields providing non-negligible amount of
SUSY breaking, there can be more model-dependent variation in the pattern
of SUSY breaking.
In the next section, we will examine a set of simple models realizing
the scenarios considered above, and evaluate the values of $R_1$ and $R_2$
in those models.

\subsection{\label{ss2.2}Soft terms}

The SUSY-breaking auxiliary components $F^T$, $F^X$, $m_{3/2}$ and
$D_A$ can generate soft SUSY-breaking masses in the visible sector
through various mediation mechanisms as described below.

A) {\it Modulus mediation}:
Since the gauge kinetic function $f_a\ni k_a T$, the $F$-term of
the GS modulus generates the gaugino masses as
\bea
M_a({\rm MM}) = F^T\partial_T \ln {\rm Re}(f_a) =
\frac{k_ag^2_a(\Lambda)}{2}F^T,
\eea
at a scale $\Lambda$ close to the Planck or string scale \cite{Moduli-mediation}.
Similarly the $T$-dependence of matter wave functions gives rise to
soft scalar masses which are generically of ${\cal O}(|F^T|^2)$:
\bea
m^2_i({\rm MM})=-|F^T|^2\partial_T\partial_{\bar{T}}
\ln(e^{-K_0/3}Z_i),
\eea
where $Z_i$ is the K\"ahler metric of the matter superfield $\phi_i$.

B) {\it Gauge mediation}:
Soft masses can receive a gauge-mediated contribution if there exist
gauge-charged messenger fields which couple to $X$
\cite{gauge-mediation-classic,Gauge-mediation}.
Indeed, in most of the known potentially (semi)realistic string models
with $U(1)_A$, there exist exotic matter fields $\Phi$, $\Phi^c$
which are vector-like under the SM gauge group and become massive through
the Yukawa coupling
\bea
\Delta W\,=\,y_\Phi X\Phi\Phi^c.
\eea
Then, gaugino masses are generated at the messenger scale
$\Lambda_\Phi=\lambda_\Phi \langle X\rangle$ according to
\bea
M_a({\rm GM}) = -\frac{N_\Phi g^2_a(\Lambda_\Phi)}{16\pi^2}
\frac{F^X}{X},
\eea
where $N_\Phi$ is the number of $\Phi+\Phi^c$ which are assumed to form
$5+\bar 5$ of $SU(5)$.
Gauge mediation induces soft scalar masses also, giving
\bea
m^2_{i}({\rm GM})={\cal O}\left(\left(\frac{1}{8\pi^2}\frac{F^X}{X}
\right)^2\right).
\eea
An interesting feature of the gauge mediation in models with $U(1)_A$ is
the contribution to scalar masses originating from $D_A$.
Because $\Phi\Phi^c$ carries a $U(1)_A$ charge $q_\Phi+q_{\Phi^c}=1$
(note that we use the convention $q_X=-1$), the supertrace of the
messenger mass matrix is non-vanishing due to the $D$-term contribution.
As a result, gauge-mediated soft scalar masses contain the piece
\cite{Extra-gauge-mediation}
\bea
\delta m^2_{i}({\rm GM})\simeq \sum_a C^a_{i} g^4_a
\frac{N_\Phi\ln(\Lambda/\Lambda_\Phi)}{(8\pi^2)^2}g^2_AD_A,
\eea
where $C^a_{i}$ is the quadratic Casimir of $\phi_i$.
This piece can be important for matter fields with $q_i=0$ when
$g_A^2D_A\gtrsim |F^X/X|^2$.

C) {\it Anomaly mediation}:
Supergravity always mediates SUSY breaking through the conformal anomaly
\cite{Anomaly-mediation}.
This effect can be described by introducing the supergravity conformal
compensator $C$ with an $F$-component:
\bea
{F^C}=m_{3/2}^*+\frac{1}{3}F^I\partial_I K,
\eea
which is generically of the order\footnote{
In some case such as the no-scale model, there can be a cancellation
between $m_{3/2}$ and $\frac{1}{3}F^I\partial_IK$, which would result in
$|F^C|\ll |m_{3/2}|$.
}
of $m_{3/2}$.
In this formulation, due to the super-Weyl invariance, the $C$-dependence
of physical gauge couplings and wavefunction coefficients is determined by
the renormalization group running, which results in
\bea
M_a({\rm AM}) &=& \frac{\beta_a}{g_a} F^C =
{\cal O}\left(\frac{m_{3/2}}{8\pi^2}\right),
\nonumber \\
m_{i}^2({\rm AM}) &=& -\frac{1}{4}
\frac{d\gamma_i}{d\ln \mu}\left|{F^C}\right|^2
= {\cal O}\left(\left(\frac{m_{3/2}}{8\pi^2}\right)^2\right),
\eea
where $\beta_a=d g_a/d\ln \mu$  and $\gamma_i=d\ln Z_i/d\ln\mu$ are
the gauge beta-function and the matter anomalous dimension, respectively.

D) {\it $D$-term contribution}:
In the presence of anomalous $U(1)_A$, there can be a $D$-term contribution
to the soft scalar mass for $U(1)_A$ charged matter fields:
\bea
m^2_i({\rm D}) = -q_i g^2_AD_A,
\eea
where $q_i$ denotes the $U(1)_A$ charge of the corresponding matter field.


\begin{figure}[t]
\begin{center}
\begin{minipage}{15cm}
\centerline{
{\hspace*{0cm}\epsfig{figure=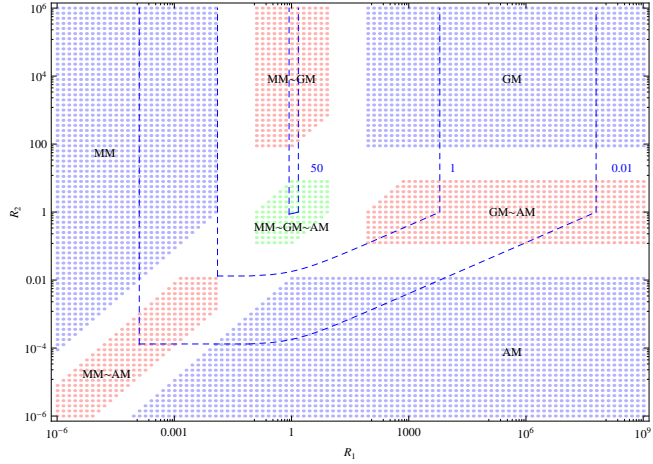,angle=0,width=8.6cm}}
}
\caption{Relative strength of each mediation.
Either modulus, gauge, or anomaly mediation dominates over the other two
in the blue region, while mixed mediations are realized in the red and
green regions.
The dashed blue line is the contour for a given value of
$m_{\rm soft}({\rm D})/\hat m_{\rm soft}$, where $\hat m_{\rm soft}$
is the biggest of $m_{\rm soft}({\rm MM,GM,AM})$.
}
\label{fig:Rel-R1}
\end{minipage}
\end{center}
\end{figure}

Let us now examine the relative importance of these mediations in
the case that
\bea
|X|^2 \sim |\xi_{FI}|\gg D_A
\eea
and $F^T$ and ${F^X}/{X}$ are at least comparable to other moduli and
matter $F$-components, respectively.
In such case, the ratios between auxiliary components are estimated as
(\ref{ratio_auxil}), and then we find the following order of magnitude relation
between different mediations:
\bea
\label{ratio4}
m_{\rm soft}({\rm MM}) : m_{\rm soft}({\rm GM}) :
m_{\rm soft}({\rm AM}) : m_{\rm soft}({\rm D})
\,\,\sim \,\,
\frac{1}{R_1}:1:\frac{1}{R_2}:\frac{8\pi^2}{\sqrt{R_1+1}}.
\eea
Note that the ratios between the modulus, gauge and $D$-term mediations are
determined essentially by a single parameter $R_1$, while the relative
importance of anomaly mediation is determined by $R_2$.
Fig. \ref{fig:Rel-R1} shows the relative strength of modulus, gauge,
and anomaly mediation as a function of $R_1$ and $R_2$.
Also shown in the figure is the contour of the ratio between
$m_{\rm soft}({\rm D})$ and the biggest of
$m_{\rm soft}({\rm MM})$, $m_{\rm soft}({\rm GM})$ and $m_{\rm soft}({\rm AM})$.
As we have noticed, both $R_1$ and $R_2$ are quite model-dependent and can have
a wide range of values.
In the subsections \ref{ss3.2} and \ref{ss3.3}, we will discuss a set of simple
models in which the SUSY breaking auxiliary components are given by
(\ref{ratio_auxil}), and therefore the soft masses obey the relation
(\ref{ratio4}).
In the subsequent subsections, we will consider other type of models which
require a separate discussion as the relation (\ref{ratio_auxil}) does
not apply.

\section{\label{s3}Stabilization of the $D$-flat direction}

In this section, we  explore with explicit examples how the $D$-flat
directions in models with $U(1)_A$ can be stabilized at a phenomenologically
viable (meta-stable) vacuum with nearly vanishing cosmological constant.
We will consider two different types of models, one in which the $D$-flatness
is achieved mostly by the cancellation between $\xi_{FI}$ and the matter field
with the largest VEV, and the other in which the $D$-flatness is achieved
mostly by the cancellation between two matter fields with opposite $U(1)_A$
charges.
For the first type of models discussed in the subsections \ref{ss3.2} and
\ref{ss3.3}, the simple relations (\ref{ratio_auxil}) and (\ref{ratio4}) are
satisfied, so the structure of mixed mediation is determined by $R_1$ and $R_2$.
On the other hand, (\ref{ratio_auxil}) and (\ref{ratio4}) do not apply
to the second type of models discussed in the subsections \ref{ss3.4} and
\ref{ss3.5}.
Our results show that various different mixed mediations can be realized within
a relatively simple class of models.

\subsection{\label{ss3.1}Uplifting potential}

Quite often, SUSY breaking by $U(1)_A$ charged fields alone cannot
give a nearly vanishing cosmological constant, and then one needs to
introduce additional SUSY breaking providing an uplifting potential
for de-Sitter or Minkowski vacuum solution.
If $U(1)_A$ charged fields all get a mass $\gg m_{3/2}$, the stabilization
would not be affected significantly by the uplifting potential.
However, in case that some $D$-flat direction has a mass $\lesssim m_{3/2}$,
the uplifting potential can play an important role for the stabilization.
This means that one needs to include the uplifting potential explicitly
in the analysis in order to draw a reliable conclusion on the stabilization.
To be specific, here we will consider a particular form of uplifting potential
which originates from a sequestered SUSY breaking sector in which SUSY is
non-linearly realized.

Then the full scalar potential of the model can be derived from the
following 4D SUGRA action:
\bea
\int d^4\theta\, C\bar{C}
\left(-3e^{-K/3}+C\bar{C} M^4\Lambda^2\bar{\Lambda}^2\right)
+ \left(\int d^2\theta\, C^3W+{\rm h.c.}\right),
\eea
where $C$ is the chiral compensator superfield introduced to encode
the SUSY breaking effects due to the auxiliary component of the SUGRA
multiplet, $K$ and $W$ are the conventional K\"ahler potential and
superpotential of the model, and $\Lambda^2\bar{\Lambda}^2$ is the
Volkov-Akulov (VA) action of the Goldstino superfield
$\Lambda^\alpha=\theta^\alpha +\frac{1}{M^2}\lambda^\alpha$, where
$\lambda^\alpha$ is the Goldstino fermion.
Here the $C$-dependence of the action is determined by the super-Weyl
invariance.
The VA action might be a low energy consequence of spontaneous SUSY
breaking at some high energy scale, e.g. a low energy realization of
the action
\bea
\int d^4\theta C\bar{C}
\left(Z\bar{Z}-\frac{Z^2\bar{Z}^2}{M^2_1}\right)
+ \left(\int d^2\theta \,C^3 M_2^2 Z +{\rm h.c.}\right),
\eea
where $Z$ is a Polonyi field, and $M_1\sim M_2 \gg m_{3/2}$ are constant
mass parameters.
Alternatively it might represent the effect of anti-brane stabilized
at the tip of warped throat in KKLT-type compactification \cite{KKLT}.
In the former case, the Goldstino scale $M$ is determined as
$M\sim M_1\sim M_2$, while in the latter case $M$ is determined
by the red-shifted tension of anti-brane.
After integrating out all auxiliary components and choosing the Einstein
frame condition for the lowest component of the compensator superfield
$C_0=e^{K/6}$, one finds that the full scalar potential is given by
\bea
V_{\rm TOT}=V_{\rm SUGRA}+V_{\rm lift},
\eea
where
\bea
V_{\rm SUGRA}=V_F +V_D= \Big(K_{I\bar J}F^I F^{J*} - 3e^K|W|^2\Big)
+ \frac{g^2_A}{2}D^2_A
\eea
is the conventional SUGRA potential and
\bea
V_{\rm lift}=M^4 e^{2K/3}
\eea
is the uplifting potential from the VA action.

\subsection{\label{ss3.2}Models with non-perturbative superpotential}

A natural source of moduli potential in string theory is non-perturbative
effect such as stringy instanton or hidden sector gaugino condensation.
So let us discuss first the stabilization of $D$-flat direction by
non-perturbative superpotential.
For simplicity, we consider the case that the GS modulus does not have
a K\"ahler mixing with other SUSY breaking moduli, and the FI term is
cancelled mostly by the $U(1)_A$ charged matter field $X$ with the
largest VEV.
Since the moduli-dependence of the matter K\"ahler metric does not change
the essential feature of stabilization, it is sufficient to consider the case
of moduli-independent matter K\"ahler metric.
Then the K\"ahler potential relevant for our discussion is given by
\bea
\label{MS-K}
K = K_0(t-\delta_{GS}V_A) + X^*e^{-2V_A}X,
\eea
where $K_0$ can take an arbitrary form.
The non-perturbative superpotential generically takes the form
\bea
\label{npw}
W_{\rm np} \propto e^{-2nT/\delta_{GS}}X^n
\eea
with an integer $n>0$ and $\delta_{GS}>0$.
Since we need to stabilize $X$ at a scale far above TeV, it is desirable
that $X$ has a flat potential in the global SUSY limit when the $D$-flat
direction is mostly $X$.
This is achieved when $n=1$, so we consider
\bea
\label{MS-W}
W = \omega_0 + A e^{-2T/\delta_{GS}}X,
\eea
where $A$ is a constant of order unity\footnote{
Note that $A$ can always be made to be of order unity through a constant
shift of $T$.
We have also chosen the normalization convention of $T$ for which
$\delta_{GS}={\cal O}(1/8\pi^2)$. Whenever we use an explicit form
of the modulus K\"ahler potential, it is defined in such field
basis.
}
and $\omega_0$ is a small constant of ${\cal O}(m_{3/2}M^2_{Pl})$.

For the K\"ahler potential (\ref{MS-K}) and the superpotential
(\ref{MS-W}), it is straightforward to see that the VEV of
$\arg(e^{-2T/\delta_{GS}}X)$ is fixed by the $F$-term potential at
\bea
\label{Phase-vacuum}
\arg(e^{-2T/\delta_{GS}}X) &=&
\left\{
\begin{array}{lll}
    \arg\left(\frac{\omega_0}{A}\right) + \pi & \quad\hbox{for}
    & R_1 < \frac{1-\xi_{FI}}{2-|X|^2}, \\
    \arg\left(\frac{\omega_0}{A}\right)       & \quad\hbox{for}
    & R_1 > \frac{1-\xi_{FI}}{2-|X|^2}, \\
\end{array}
\right.
\eea
where $R_1=-\delta_{GS}K_0^{\prime\prime}/2K_0^{\prime}$ as defined in
(\ref{r1r2}) and $\xi_{FI}=\delta_{GS}K_0^\prime/2$.
Meanwhile, one combination of ${\rm Im}(T)$ and $\arg(X)$ remains
unfixed, and is absorbed into the $U(1)_A$ gauge boson.
After replacing $\arg(e^{-2T/\delta_{GS}}X)$ with its VEV, the remaining
$t=T+T^*$ and $|X|$ can be fixed by the stationary condition
\bea
\label{STA}
\partial_{t,|X|}\left(V_{\rm SUGRA}+V_{\rm lift}\right)=0\eea
under the constraint \bea \langle V_{\rm SUGRA}+V_{\rm lift}\rangle
=0.\eea

To proceed, let us first consider the case with
\bea
R_1\ll 1.
\eea
We find that in this case the uplifting potential can be treated as a small
perturbation, so one can start with a solution of
\bea
\label{STA1}
\partial_{t,|X|}V_{\rm SUGRA}=0.
\eea
Then the stationary conditions for $V_{\rm SUGRA}$ give rise to
\bea
\label{Ext-cond}
g^2_A D_A &=&
-\frac{1}{\frac{\delta_{GS}}{2}K^{\prime\prime}_0}
\left(V^\prime_F + (\partial_t\ln g^2_A)V_D \right),
\nonumber \\
\dot V_F  &=&
\frac{2|X|^2}{\frac{\delta_{GS}}{2}K^{\prime\prime}_0}
\left(V^\prime_F + (\partial_t\ln g^2_A)V_D \right),
\eea
where the prime and dot denote the derivatives with respect to $t$ and
$\ln|X|$, respectively.
The second relation above determines how the $D$-flat direction is fixed
by the $F$-term potential.
Neglecting small corrections of ${\cal O}(m^4_{3/2}/M^4_V)$, the condition
for the solution of (\ref{Ext-cond}) to be a (local) minimum of $V_{\rm SUGRA}$
reads
\bea
\label{Min-cond}
\ddot V_F + 4 \frac{|X|^4}{\frac{\delta^2_{GS}}{4}
K^{\prime\prime 2}_0} \left( V^{\prime\prime}_F
- \frac{K^{\prime\prime\prime}_0}{K^{\prime\prime}_0} V^\prime_F
\right) - 4\frac{|X|^2}{\frac{\delta_{GS}}{2}K^{\prime\prime}_0}
\dot V^\prime_F > 0.
\eea
Note that a supersymmetric solution of (\ref{Ext-cond})  leads to
$\dot V_F=V^\prime_F=0$ and $D_A=0$.

It turns out that, in the limit $R_1\ll 1$, the equations (\ref{Ext-cond})
have a unique solution which is supersymmetric and automatically fulfills
the (meta)stability condition (\ref{Min-cond}) regardless of the form of $K_0$.
The minimum of $V_{\rm SUGRA}$ is given by
\bea
\label{SSF}
|X|^2 &=&
-\frac{\delta_{GS}}{2}K_0^\prime,
\nonumber \\
\left|\frac{A}{\omega_0}\right|e^{-t/\delta_{GS}} &=&
\frac{|X|}{1-|X|^2}\,\simeq\, |X|,
\eea
and the vacuum solution of $V_{\rm TOT}$ can be obtained by taking into account
the small shift from this solution induced by $V_{\rm lift}$.
It is then easy to find that the vacuum solution of $V_{\rm TOT}$ gives
\bea
F^T &\simeq&
\frac{\delta_{GS}}{2}\frac{3\partial_t\ln V_{\rm lift}}{K_0^\prime}m_{3/2}
\,=\, \delta_{GS} m_{3/2}\,=\,
{\cal O}\left(\frac{m_{3/2}}{8\pi^2}\right),
\nonumber \\
\frac{F^X}{X} &\simeq&
\left(\frac{K_0^{\prime\prime}}{K_0^\prime}\right){F^T}\,\simeq\,
-2R_1 m_{3/2},
\eea
where we have used the sequestered uplifting potential $V_{\rm lift}=M^4 e^{2K/3}$.
The above result also leads to $R_2\simeq 2R_1$.
The $D$-flat direction, which is mostly $T$ in this case, acquires a mass of
$\sim m_{3/2}\ln(M_{Pl}/m_{3/2})\gg m_{3/2}$, and this is the reason why
$F^T\ll m_{3/2}$.

Another limit for which the analysis is straightforward is the case with
\bea
R_1 \gg 1.
\eea
Since the non-perturbative superpotential stabilizes the GS modulus at
$t/\delta_{GS}\gg 1$ in the field basis with $A={\cal O}(1)$, it is plausible
to assume that the modulus K\"ahler potential satisfies
\bea
\label{assume}
\left| \left(\frac{\delta_{GS}}{2} \right)^{k-2}
\frac{\partial^k_t K_0}{K^{\prime\prime}_0} \right|
\lesssim \left(\frac{\delta_{GS}}{t}\right)^{k-2} \ll 1
\qquad
(k\geq 3)
\eea
at the stationary point of $V_{\rm SUGRA}$.
We then find that $V_{\rm SUGRA}$ can have a SUSY breaking minimum at
\bea
\label{SBF}
|X|^2 &=&
-\frac{\delta_{GS}}{2}K_0^\prime +
{\cal O}\left(\frac{m^2_{3/2}M^2_{Pl}}{M^2_V}\right),
\nonumber \\
\left|\frac{A}{\omega_0}\right| e^{-t/\delta_{GS}} &=&
\left({R_1} + {\cal O}\left(\frac{M^2_V}{M^2_{Pl}}
\right)\right)|X|,
\eea
for $M_{Pl}\gg M_V \gg (m_{3/2}M_{Pl})^{1/2}$.
It is straightforward to show that the above field configuration satisfies
the stability condition (\ref{Min-cond}) and leads to
\bea
\label{V-at-SB}
\langle V_{\rm SUGRA}\rangle =
e^K|\omega_0|^2 \left( -3 + {R_1}\frac{M^2_{GS}}{M^2_{Pl}}
+ {\cal O}\left(\frac{M^2_{GS}}{M^2_{Pl}}\right) \right),
\eea
where the second term in the brackets is the contribution from $F^X$,
and thus $R_2\simeq R_1$ at the minimum of $V_{\rm SUGRA}$.
The above form of the vacuum energy density suggests that one can get
a de-Sitter or Minkowski minimum {\it without} introducing an uplifting
potential, if
\bea
R_1 ={\cal O}\left(\frac{M^2_{\rm Pl}}{M^2_{GS}}\right)=
{\cal O}\left(\frac{1}{\delta_{GS}^2}\right).
\eea
In this case, $X$ (approximately) corresponds to the $D$-flat direction,
and its scalar component acquires a mass of ${\cal O}(m_{3/2}M_{Pl}/M_{GS})$,
while its fermionic component corresponds to the Goldstino absorbed
into the gravitino.

Unlike the case with $R_1\ll 1$ or $R_1\gg 1$, the analysis of the vacuum
solution of $V_{\rm TOT}$ for $R_1={\cal O}(1)$ is quite nontrivial.
In such case, we need more model-dependent and detailed analysis to make
sure that there exists a (local) minimum of $V_{\rm TOT}$ with vanishing
cosmological constant.

In the following, to examine explicitly the stabilization of $T$ and
$X$, we consider two different forms of the modulus K\"ahler potential
\bea
\begin{array}{lll}
K^{(I)}_0 &=& -n_0\ln(t-\delta_{GS}V_A), \\
K^{(II)}_0 &=&
\frac{1}{2}K_0^{\prime\prime}(t_0)(t-t_0-\delta_{GS}V_A)^2,
\end{array}
\eea
with the superpotential and the uplifting potential given by
\bea
W &=& \omega_0 + AXe^{-2T/\delta_{GS}},
\nonumber \\
V_{\rm lift} &=& M^4e^{2K/3}. \eea In $K_0^{(I)}$, $T$ might
correspond to a dilaton in the weak coupling limit or a volume
modulus in the large volume limit.
On the other hand, $K_0^{(II)}$ assumes that there exists a point
in the moduli space where the FI term vanishes
\cite{uranga,Stability wall,GS-blowing-up},
\bea
\xi_{FI}(t_0)=\frac{1}{2}\delta_{GS}K_0^\prime(t=t_0)=0,
\eea
and then the modulus K\"ahler potential is expanded around $t=t_0$.
Using the field redefinition $T\rightarrow \alpha T+\beta$ for real
$\alpha$ and $\beta$, we can choose the convention
\bea
|A|=1, \quad \delta_{GS}=\frac{1}{2\pi^2}.
\eea
Generically $K_0^{\prime\prime}(t_0)$ is expected to be of order unity,
however $\omega_0$ is required to be exponentially small to realize low
energy SUSY.
To be specific, we choose
\bea
K^{\prime\prime}(t_0)=1, \quad |\omega_0| = e^{-38},
\eea
and examine the following four examples:
\bea
\begin{tabular}{lcll}
Model I   & : & $K_0 =K_0^{(I)}$ & with $n_0=1$,    \\
Model II  & : & $K_0=K_0^{(II)}$ & with $t_0=2.6$,  \\
Model III & : & $K_0=K_0^{(II)}$ & with $t_0=2.1$,  \\
Model IV  & : & $K_0=K_0^{(II)}$ & with $t_0=1.9$.  \\
\end{tabular}
\eea

\begin{figure}[t]
\begin{center}
\begin{minipage}{15cm}
\centerline{
{\hspace*{0cm}\epsfig{figure=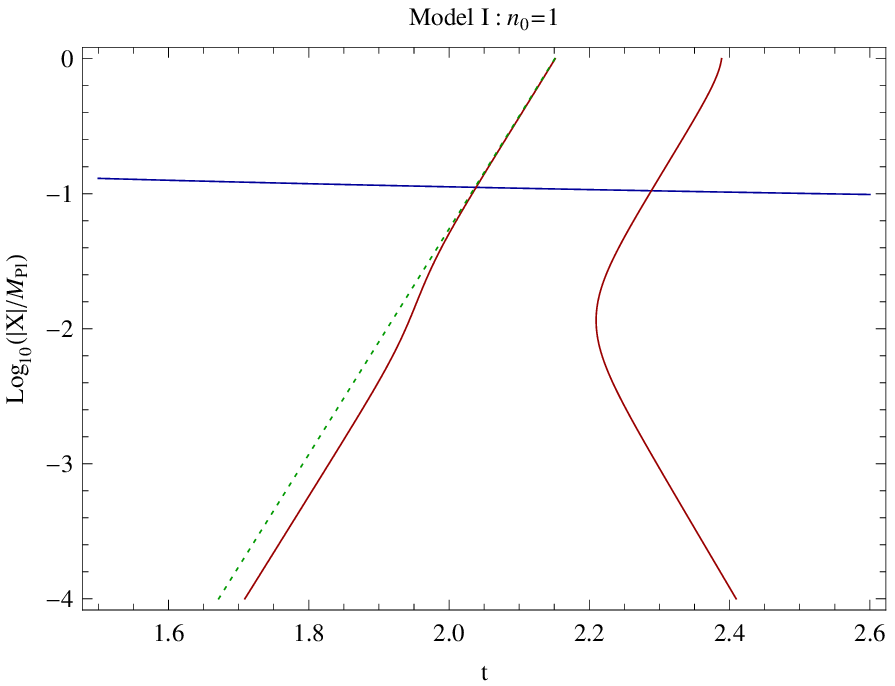,angle=0,width=7.3cm}}
{\hspace*{.2cm}\epsfig{figure=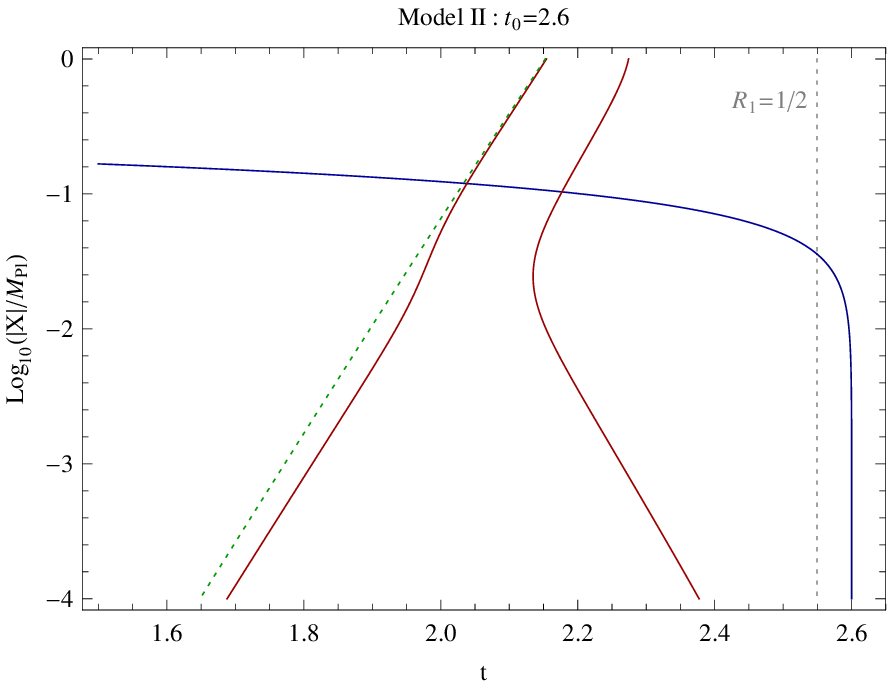,angle=0,width=7.3cm}}
}
\vspace{0.8cm}
\centerline{
{\hspace*{0cm}\epsfig{figure=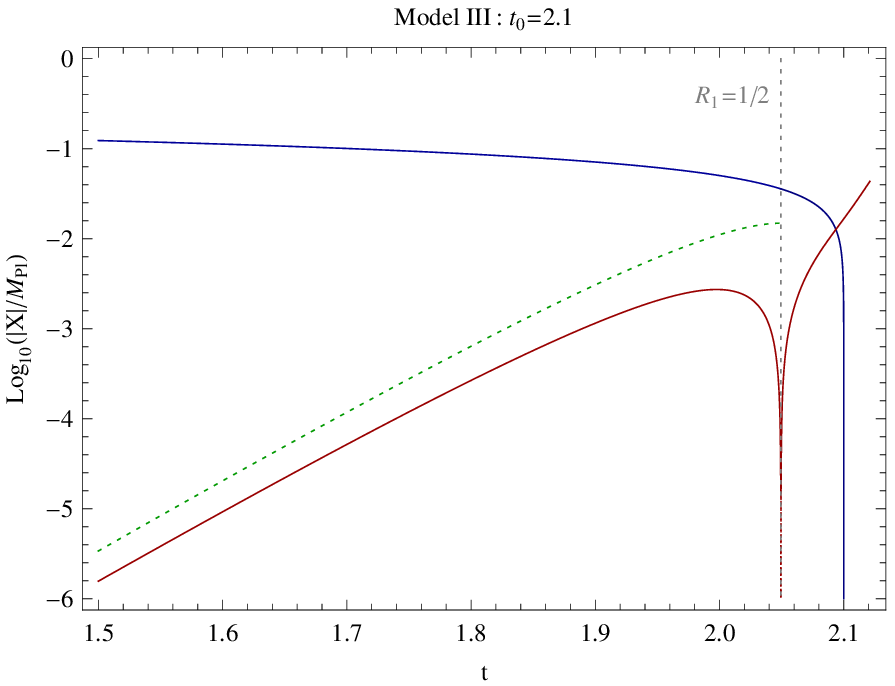,angle=0,width=7.3cm}}
{\hspace*{.2cm}\epsfig{figure=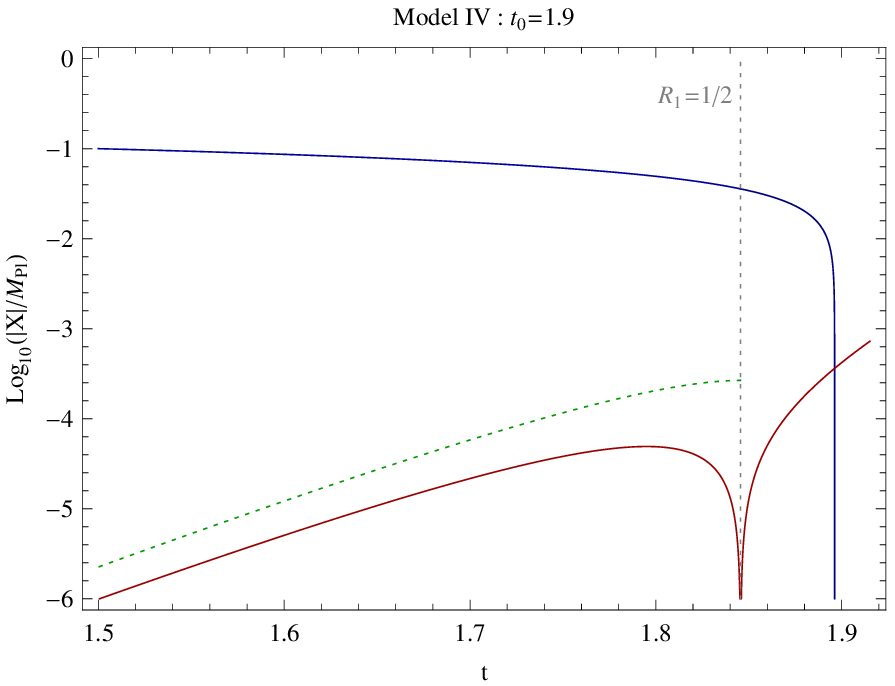,angle=0,width=7.3cm}} }
\caption{ Stabilization of the GS modulus $t=T+T^*$ and the chiral
matter $\ln |X|$ in each model.
In each figure, the blue curve corresponds to the $D$-flat condition,
while the red curve represents the stationary condition,
$(\partial_t D_A)\partial_{|X|}V_{\rm TOT}-(\partial_{|X|}D_A)
\partial_t V_{\rm TOT}=0$.
We also plot the dashed green curve along which $F^T$ vanishes.
For Model II, III and IV, $R_1=1/2$ on the vertical dashed line.}
\label{fig:Stabilization}
\end{minipage}
\end{center}
\end{figure}

Fig. \ref{fig:Stabilization} shows how a minimum of
$V_{\rm TOT}=V_{\rm SUGRA}+V_{\rm lift}$ with vanishing cosmological
constant is developed in these examples.
In each figure, the blue curve corresponds to the $D$-flat condition,
while the red curve represents the stationary condition\footnote{
This stationary condition corresponds to the second relation
of (\ref{Ext-cond}) for Model IV.
On the other hand, for other models, one needs to replace $V_F$
with $V_F+V_{\rm lift}$ in (\ref{Ext-cond}) since the uplifting potential
is needed to cancel the cosmological constant.
},
$(\partial_t D_A)\partial_{|X|}V_{\rm TOT}-(\partial_{|X|}D_A)
\partial_t V_{\rm TOT}=0$.
Hence the point where the two curves intersect corresponds to a stationary
solution of $V_{\rm TOT}$.
In Model I and II, we have two intersecting  points, and the point with
smaller $t$ is a (local) minimum, while the other point is a saddle point.
The vacuum solution can be considered as a small deviation from the
supersymmetric solution which is the intersecting point of the blue and
dashed green curves in the figure, and this small deviation is due to
$V_{\rm lift}$.
On the other hand, for Model III and IV, the blue and red curves have
a unique intersecting point which is a local minimum of the potential.
Here we have tuned the size of $V_{\rm lift}$ to make the vacuum solution
have a vanishing cosmological constant.
We then find that the VEVs of
\bea
\frac{|X|^2}{M_{Pl}^2}\simeq -\frac{\xi_{FI}}{M_{Pl}^2},\quad
R_1=-\frac{\delta_{GS}}{2}\frac{K_0^{\prime\prime}}{K_0^\prime},\quad
|R_2|=\left|\frac{D_XW}{X W}\right|
\eea
are given by
\bea
\begin{tabular}{lccc}
Model I    &\,\, $1.6\times 10^{-2}$,
& $1.6\times 10^{-2}$, & $3.1 \times 10^{-2}$  \\
Model II   &\,\, $1.5 \times 10^{-2}$,
& $4.2\times 10^{-2}$, &  $8.4\times 10^{-2}$ \\
Model III  &\,\, $1.6 \times 10^{-4}$,
& $4$, & $3.8$ \\
Model IV   &\,\, $1.4\times 10^{-7}$,
& $4.7\times 10^3,$ & $4.7\times 10^3$ \\
\end{tabular}
\eea
and therefore the ratios
\bea
F^T: \frac{F^X}{X} : m^*_{3/2}:
g_A\sqrt{D_A} \, \simeq  \, \frac{\delta_{GS}}{2R_1}:\, -1\,:
\,\frac{1}{R_2}\, :\frac{1}{\sqrt{R_1+1}}
\eea
are determined as
\bea
\begin{tabular}{lccccccc}
Model I   &\,\,  1.6               & : & $-1$ & : & 32 & : & 1      \\
Model II  &\,\,  0.6               & : & $-1$ & : & 12 & : & 1      \\
Model III &\,\,  $6\times 10^{-3}$ & : & $-1$ & : & 0.26 & : & 0.45 \\
Model IV  &\,\,  $5\times 10^{-6}$ & : & $-1$ & : & $2\times10^{-4}$
& : & $1.5\times 10^{-2}$ \\
\end{tabular}
\eea
With the above results, the ratios between the modulus, gauge,
anomaly and $D$-term mediated soft masses in each model can be read
off from the following order of magnitude relation
\bea
m_{\rm soft}({\rm MM}) : m_{\rm soft}({\rm GM})
: m_{\rm soft}({\rm AM}) :
m_{\rm soft}({\rm D})  \, \sim \, \frac{1}{R_1} \,:\, 1 \,:\,
\frac{1}{R_2} \,:\, \frac{8\pi^2}{\sqrt{R_1+1}},
\eea
where we have assumed that there exist gauge-charged messengers $\Phi+\Phi^c$
which become massive through the coupling to $X$:
\bea
\Delta W\,=\,
y_\Phi X\Phi\Phi^c.
\eea

Our results indicate that the relative importance of each mediation
is quite model-dependent, particularly on the form of the modulus
K\"ahler potential. Models I and II realize mixed modulus-anomaly-$D$-term
mediation, which corresponds to the mirage mediation
\cite{KKLT-mirage-med,Mirage-mediation-phen,Little-hierarchy-in-mirage-mediation}
with additional $D$-term contribution \cite{KKLT-U1A}.
On the other hand, in Model III, gaugino masses are determined by mixed
modulus-gauge-anomaly mediation, while scalar masses are dominated by
the $D$-term contribution which is one or two orders of magnitudes heavier
than gaugino masses.
Finally, soft masses in Model IV are determined by mixed gauge-$D$-term mediation.
So various different mixed mediations can be realized even within the simple
class of models discussed in this subsection.

Since the $U(1)_A$ vector superfield acquires a large supersymmetric mass,
while leaving the $D$-flat combination of $T$ and $\ln X$ light,
we can study the low energy dynamics of model with an effective supergravity
in which the massive $U(1)_A$ vector superfield is integrated out.
In the limit $R_1\ll 1$ or $R_1\gg 1$, the massive and light degrees of freedom
can be easily identified, and therefore it is straightforward to derive
the effective theory of light fields in this limit.
For instance, the relations
\bea
\frac{M_V^2}{2g_A^2}\,\simeq\, \frac{1}{4}\delta_{GS}^2
K_0^{\prime\prime}+|X|^2, \quad
R_1\,\simeq\,
\frac{\delta_{GS}^2}{4}\frac{K_0^{\prime\prime}}{|X|^2}
\eea
suggest that the Goldstone superfield absorbed into the massive $U(1)_A$
vector superfield is mostly $\ln X$ in the limit $R_1\ll 1$, while
it is mostly $T$ in the opposite limit $R_1\gg 1$.

Since it provides an efficient way to understand our results, let us
derive the effective theory explicitly in the limit $R_1\ll 1$ or
$R_1\gg 1$.
When $R_1\ll 1$, the massive vector superfield and the light $D$-flat direction
are identified as
\bea
V_H = V_A -\ln\left(\frac{|X|}{M_{Pl}}\right),
\quad \tilde T = T-\frac{1}{2}\delta_{GS}
\ln\left(\frac{X}{M_{Pl}}\right),
\eea
both of which are invariant under $U(1)_A$.
Then it is convenient to introduce the $U(1)_A$-invariant combination of
matter fields
\bea
\tilde \phi_i =
\left(\frac{X}{M_{Pl}}\right)^{q_i}\phi_i,
\eea
and rewrite the K\"ahler potential and superpotential in terms of $V_H$,
$\tilde T$ and $\tilde \phi_i$:
\bea
\label{full1}
K &=& K_0
(t-\delta_{GS}V_A)+X^*e^{-2V_A}X+ \phi_i^*e^{2q_i V_A}\phi_i
\nonumber \\
&=& K_0(\tilde t-\delta_{GS}V_H)+e^{-2V_H}
+ \tilde\phi_i^*e^{2q_i V_H}\tilde\phi_i,
\nonumber \\
W &=& \omega_0 +AXe^{-2T/\delta_{GS}}
+ \lambda_{ijk}X^{q_i+q_j+q_k}\phi_i\phi_j\phi_k
\nonumber \\
&=& \omega_0 +A e^{-2\tilde T/\delta_{GS}}
+\lambda_{ijk}\tilde\phi_i\tilde\phi_j\tilde\phi_k,
\eea
where $t=T+T^*$ and $\tilde t =\tilde T+\tilde T^*$.
Now the massive vector superfield $V_H$ can be integrated out by solving
the equation of motion
\bea
\label{eqm}
\partial_{V_H}K=0,
\eea
whose solution is given by
\bea
\frac{2}{\delta_{GS}}e^{-2V_H} = K_0^\prime(\tilde t)
+{\cal O}(\delta_{GS}).
\eea
Inserting this solution into the K\"ahler potential, we find the effective
K\"ahler potential of $\tilde T$ and $\tilde \phi_i$ ($\neq\tilde{X}$) is
given by
\bea
K_{\rm eff} &=& K_0 (\tilde t) +{\cal O}(\delta_{GS})+
\sum_{i\neq X}\left(\frac{\delta_{GS}}{2}\left(
K_0^\prime(\tilde t)+{\cal O}(\delta_{GS})\right)
\right)^{-q_i}|\tilde\phi_i|^2,
\eea
from which all the low energy consequences of (\ref{full1}),
including the SUSY breaking auxiliary components of light fields and
the soft terms of visible sector fields, can be derived in the
approximation in which the subleading corrections suppressed by
$\delta_{GS}$ are ignored.

On the other hand, in the other limit $R_1\gg 1$, the massive vector
superfield and the light matter fields
(including the $D$-flat direction $\tilde{X}$) are given by
\bea
V_H = V_A -\frac{1}{\delta_{GS}}(t-t_0), \quad \tilde \phi_i = \phi_i
e^{2q_i(T-T_0)/\delta_{GS}},
\eea
where $t_0$ is the modulus value for which the FI term vanishes:
\bea
K_0^\prime (t=t_0)=0.
\eea
We then have
\bea
K &=& K_0 (t-\delta_{GS}V_A)+ \sum_{I=X,i}\phi_I^*e^{2q_I V_A}\phi_I
\nonumber \\
&=&
K_0(t_0-\delta_{GS}V_H)
+ \sum_{I=X,i}\tilde\phi_I^*e^{2q_I V_H}\tilde\phi_I,
\nonumber \\
W &=& \omega_0 + AXe^{-2T/\delta_{GS}}
+ \lambda_{ijk}X^{q_i+q_j+q_k}\phi_i\phi_j\phi_k
\nonumber \\
&=& \omega_0 +A e^{-2T_0/\delta_{GS}}\tilde X
+ \lambda_{ijk}\tilde{X}^{q_i+q_j+q_k}\tilde\phi_i\tilde\phi_j\tilde\phi_k,
\eea
for which the solution of the equation of motion (\ref{eqm}) is given by
\bea
\label{vh1}
V_H =-\frac{\sum_I q_I \tilde \phi_I^*\tilde\phi_I}{2M_{GS}^2}
+{\cal O}\left(\frac{|\tilde\phi|^4}{M_{GS}^4}\right),
\eea
where $M_{GS}^2=\delta_{GS}^2K_0^{\prime\prime}(t=t_0)/4$.
The resulting effective K\"ahler potential and superpotential are given by
\bea
\label{eff1}
K_{\rm eff} &=& K_0(t_0)+\sum_{I=X,i} |\tilde\phi_I|^2
-\sum_{IJ=X,i}\frac{q_Iq_J}{2M_{GS}^2}|\tilde\phi_I|^2|\tilde\phi_J|^2
+{\cal O}\left(\frac{|\tilde\phi|^6}{M_{GS}^4}\right),
\nonumber \\
W_{\rm eff} &=& \omega_0 + Ae^{-2T_0/\delta_{GS}}\tilde{X}
+ \lambda_{ijk}\tilde{X}^{q_i+q_j+q_k}
\tilde\phi_i\tilde\phi_j\tilde\phi_k.
\eea
One is thus led to the low energy theory where $\tilde X$ acts as a Polonyi
field and is stabilized by the same way as in sweet-spot SUSY
\cite{Sweet-spot-SUSY}.
In that scenario, the Higgs sector feels SUSY breaking also through direct
interactions with the Polonyi field, in addition to gauge-mediated one.
But, the higher dimensional operator
$q_i|\tilde X|^2|\tilde \phi_i|^2/M^2_{GS}$ in the above effective K\"ahler
potential can transmit SUSY breaking not only to the Higgs fields but also
to the sfermions if they are charged under $U(1)_A$.

\subsection{\label{ss3.3}Models with radiative stabilization}

In some case, non-perturbative superpotential of the GS modulus might not
be available because either $\delta_{GS}<0$, so it is forbidden by $U(1)_A$,
or the corresponding instanton amplitude is vanishing due to the zero mode
structure.
Even in such case, if the K\"ahler potential admits a limit with $\xi_{FI}=0$
or more generally a limit with $\xi_{FI}$ far below $M_{Pl}^2$, for which
the $D$-flat direction is described by $X$, the scalar potential of $D$-flat
direction can receive a field-theoretic radiative correction which can fix
the VEV of $X$ at a proper value.

As an example of such model, let us consider
\bea
K &=& K_0(t-\delta_{GS}V_A)+\sum_i \phi_i^*e^{2q_iV_A}\phi_i,
\nonumber \\
W &=& \omega_0 + y_\Phi X\Phi\Phi^c,
\eea
where
$\phi_i=\{X,\Phi,\Phi^c\}$ for the exotic matter fields $\Phi+\Phi^c$ which form
$5+\bar{5}$ of $SU(5)$ with the $U(1)_A$ charges satisfying $q_\Phi+q_{\Phi^c}=1$,
and $K_0$ is assumed to satisfy
\bea
K_0^{\prime}(t=t_0)=0.
\eea
It is straightforward to analyze this model if one uses the effective theory
in which the massive $U(1)_A$ vector superfield $V_H=V_A-(t-t_0)/\delta_{GS}$
is integrated out.
Using the procedure described in the previous section
(see Eqs. (\ref{vh1}) and (\ref{eff1})), one easily finds
\bea
\label{sol1}
V_H = \frac{|\tilde X|^2}{2M_{GS}^2}+\cdots,
\eea
and the resulting effective K\"ahler potential takes the form
\bea
K_{\rm eff}=
|\tilde X|^2-\frac{1}{2}\frac{|\tilde X|^4}{M_{GS}^2}+\cdots,
\eea
where the ellipses denote the terms involving $\Phi$, $\Phi^c$.
Then the scalar potential of $|X|$ at tree level includes
a quartic term originating from the quartic term in $K_{\rm eff}$:
\bea
(V_{\rm SUGRA}+V_{\rm lift})|_{\rm tree}=
\frac{1}{2}\frac{m_{3/2}^2}{M^2_{GS}}|\tilde X|^4+\cdots.
\eea
There are also radiative corrections to the scalar potential for $|X|$,
in particular the anomaly mediated soft scalar mass associated with the
Yukawa coupling $y_\Phi X\Phi\Phi^c$, which is given by
\bea
\Delta V = N_\Phi y^2_\Phi\Big( 5 (5N_\Phi+2)y^2_\Phi -
16g^2_3-6g^2_2-2g^2_1 \Big)
\left(\frac{m_{3/2}}{16\pi^2}\right)^2|\tilde X|^2,
\eea
where $N_\Phi$ is the number of $\Phi+\Phi^c$, and $g_a$ are the SM gauge
couplings at the scale $y_\Phi|X|$.
In order for $X$ to develop nonzero VEV, we need
\bea
5(5N_\Phi+2)y^2_\Phi < 16g^2_3+6g^2_2+2g^2_1
\eea
at the scale $y_\Phi|X|$.
Then, the VEV of $|\tilde X|$ is determined as
\bea
\langle |\tilde X|\rangle = \left(N_\Phi
y^2_\Phi\Big(16g^2_3+6g^2_2+2g^2_1-5(5N_\Phi+2)y^2_\Phi \Big)\right)^{1/2}
\frac{M_{GS}}{16\pi^2},
\eea
for which
\bea
R_1 =
\frac{M_{GS}^2}{|\tilde X|^2} =
\frac{(16\pi^2)^2}{N_\Phi y^2_\Phi( 16g^2_3+6g^2_2+2g^2_1 - 5
(5N_\Phi+2)y^2_\Phi)}.
\eea

In fact, this model possesses an anomalous global Peccei-Quinn (PQ) symmetry
\bea
U(1)_{PQ} : \quad \tilde\phi_i \rightarrow e^{q_i\alpha}
\tilde \phi_i
\eea
which can solve the strong CP problem through the axion mechanism
\cite{PQ-mechanism,review-axion}.
Then the phase degree of freedom of $\tilde X$ can be identified as the QCD axion
with a decay constant $v_{PQ}= \langle |\tilde X|\rangle$ which is constrained
as $10^9{\rm GeV} \lesssim v_{PQ} \lesssim 10^{12}{\rm GeV}$.
The axion scale of the model can be in this range if
\bea
10^{-5}\lesssim y_\Phi \lesssim 10^{-2}.
\eea
On the other hand, for this range of $y_\Phi$, we have
\bea
R_1\,\gg\, (8\pi^2)^2,\quad
R_2=1+\left(\frac{\partial_XW}{X^*W}\right)^* \,\simeq\,1.
\eea
Applying this result to
\bea
F^T: \frac{F^X}{X} : m^*_{3/2}: g_A\sqrt{D_A} &\simeq &
\frac{\delta_{GS}}{2R_1}:\, -1\,: \,\frac{1}{R_2}\,
:\frac{1}{\sqrt{R_1+1}},
\eea
we find the modulus, gauge, anomaly and $D$-term mediated soft masses are
estimated as
\bea
m_{\rm soft}({\rm GM}) \,\sim \, m_{\rm soft}({\rm AM})\, \gg\, \,
m_{\rm soft}({\rm D})\,\gg \, m_{\rm soft}({\rm MM}),
\eea
and therefore soft masses in this model are determined by mixed gauge-anomaly
mediation \cite{anomaly-gauge,axion-in-anomaly-mediation}.
The radial scalar and fermion components of $\tilde X$ correspond to the saxion
$s$ and the axino $\tilde a$, respectively.
Their masses are given by
\bea
m_s \,\sim\, \frac{m_{3/2}}{\sqrt{R_1}},\quad m_{\tilde a} \,\sim
\,\frac{m_{3/2}}{R_1},
\eea
which suggest that the axino is the LSP in this model.

\subsection{\label{ss3.4}Models with non-renormalizable superpotential}

In this subsection, we consider a model with additional $U(1)_A$-charged
matter $Y$ with $q_Y>0$, which can cancel the $D$-term contribution from $X$.
Then there are two $D$-flat directions in the model, parameterized by
$Xe^{-2T/\delta_{GS}}$ and $X^{q_Y}Y$.
The superpotential is allowed to contain these two $U(1)_A$-invariant
holomorphic operators.
Here we discuss only the case when $q_Y=-3q_X=3$, and the superpotential
contains $X^3Y$, but no nonperturbative term involving
$Xe^{-2T/\delta_{GS}}$.

Then the K\"ahler potential and superpotential of the model are given by
\bea
K &=& K_0(t-\delta_{GS}V_A)+ X^*e^{-2V_A}X + Y^*e^{6V_A}Y,
\nonumber \\
W &=& \omega_0 + \lambda \frac{X^3 Y}{M_{Pl}}+y_\Phi X\Phi\Phi^c,
\eea
where the last term in the superpotential is not crucial for the stabilization
of $D$-flat directions, but is introduced for the gauge mediation of SUSY
breaking by $F^X$.
A key assumption on the model is that the K\"ahler potential of the GS modulus
admits a point with vanishing FI term:
\bea
K_0^\prime (t=t_0)=0.
\eea
We also assume for simplicity that the matter K\"ahler metric of $X$ and $Y$
are independent of the GS modulus, however our results equally apply
to the case with moduli-dependent matter K\"ahler metric.

Now the  $U(1)_A$ $D$-term is given by
\bea
g_A^2 D_A= \xi_{FI}+|X|^2-3|Y|^2,
\eea
where $\xi_{FI}=\delta_{GS}K_0^\prime/2$.
In section 2 and also in the previous subsections, we were focusing on the case
that $D$-flatness is achieved through the cancellation between $\xi_{FI}$ and
$|X|^2$, yielding $|\xi_{FI}|\simeq |X|^2 \gg g_A^2 D_A$.
Note that $X$ was defined as the $U(1)_A$-charged matter field with the largest VEV.
In fact, this model realizes a different scenario with
\bea
|X|^2 \sim |Y|^2 \gg \xi_{FI},
\eea
and as a consequence does not obey the relations in (\ref{FD}) except for the
last one.

For the above K\"ahler potential and superpotential, one of the stationary
conditions for scalar potential takes the form
\bea
\partial_t \left(V_{\rm SUGRA}+V_{\rm lift}\right) &=& \left(
V_F + \frac{2}{3}V_{\rm lift} + \left(2 -\frac{K^\prime_0
K^{\prime\prime\prime}_0}{K^{\prime\prime2}_0} \right)e^K|W|^2
\right) K^\prime_0
\nonumber \\
&& +\, \frac{\delta_{GS}}{2}K^{\prime\prime}_0g^2_AD_A
+ (\partial_t \ln g^2_A)\frac{g_A^2}{2}D_A^2 = 0,
\eea
which is satisfied by the $D$-flat direction given by
\bea
t=t_0, \quad |X|^2=3|Y|^2.
\eea
In the region with $|X|^2\ll M^2_{GS}=\delta_{GS}^2K_0^{\prime\prime}/4$,
the scalar potential along this $D$-flat direction is written as
\bea
V_{\rm TOT} \simeq e^K \left( 4|\lambda|^2 |X|^6
- \frac{2}{\sqrt3}|\lambda\omega_0||X|^4
- 3|\omega_0|^2 \right),
\eea
for ${\rm Arg}(X^3Y)$ fixed at ${\rm Arg}(\lambda^*\omega_0)+\pi$ by the minimization
condition.
Here we have set $M_{Pl}=1$, and neglected small corrections suppressed by
$|X|^2/M^2_{Pl}$.
This determines the VEVs of $X$ and $Y$ as
\bea
|X|^2 \,=\, 3|Y|^2 \,\simeq\,
\frac{\sqrt{3}}{9|\lambda|}m_{3/2}M_{Pl}.
\eea
The resulting SUSY breaking auxiliary components are given by
\bea
\frac{F^X}{X} \,=\, \frac{F^Y}{Y} \,\simeq\, -\frac{2}{3}m_{3/2},
\quad F^T \,=\, D_A \,=\, 0,
\eea
so soft masses are determined by mixed gauge-anomaly mediation:
\bea
m_{\rm soft}({\rm GM}) \,\sim\, m_{\rm soft}({\rm AM}) \,\gg\,
m_{\rm soft}({\rm MM}), \, m_{\rm soft}({\rm D}).
\eea
For $X$ stabilized at $|X|\ll M_{GS}$, the longitudinal component of
the massive $U(1)_A$ vector superfield comes mostly from $T$.
The non-renormalizable superpotential term provides masses to $X$ and $Y$,
which include two radial scalars $s_{1,2}$, the massive angular scalar $a_h$,
and two fermions $\tilde a_{1,2}$:
\bea
&&
m_{s_1}\simeq \frac{1}{3}m_{3/2}, \quad
m_{s_2}\simeq \frac{\sqrt3}{3}m_{3/2}, \quad
m_{a_h}\simeq \frac{\sqrt6}{3}m_{3/2},
\nonumber \\
&&
m_{\tilde a_1} \simeq 0.1 m_{3/2}, \quad
m_{\tilde a_2} \simeq 0.8 m_{3/2}.
\eea
One combination of ${\rm Arg}(X)$ and ${\rm Arg}(Y)$ remains unfixed.

We note that the above model possesses a PQ symmetry:
\bea
U(1)_{PQ}:\quad \phi_i \rightarrow e^{q_i\alpha} \phi_i
\eea
which is spontaneously broken by the VEVs of $X$ and $Y$.
The corresponding axion scale $v_{PQ}$ can take naturally an intermediate
scale value
\bea
v_{PQ}\,\sim\, \sqrt{m_{3/2}M_{Pl}}\,\sim\, 10^{11}{\rm GeV},
\eea
when $m_{3/2}\sim 10^4$GeV for which
$m_{\rm soft}\sim \frac{g^2}{8\pi^2} m_{3/2}$ has a weak scale size.

\subsection{\label{ss3.5}KKLT with $D$-term stabilization}

In this subsection, we discuss a class of models in which multiple number
of moduli, including the GS modulus, participate in SUSY breaking.
As a concrete example, we consider a variant of KKLT scenario with multiple
number of K\"ahler moduli, in which the visible sector K\"ahler modulus $T$
is stabilized by the $D$-term potential of $U(1)_A$,
while the other K\"ahler moduli are stabilized by nonperturbative
superpotential as in the original KKLT scenario \cite{KKLT}.

The motivation for this variant is the observation that instantons
wrapping the visible sector 4-cycle in KKLT setup have SM-charged
zero modes of chiral fermions, and as a result the corresponding
nonperturbative superpotential of $T$ should involve a gauge
invariant product of SM-charged chiral matter superfields
\cite{Blumenhagen:2007sm}. This would effectively make the
nonperturbative superpotential of $T$ vanish, and then one needs
other mechanism to stabilize the visible sector K\"ahler modulus. As
we will see, if the model contains an anomalous $U(1)_A$ with $T$
being the GS modulus, and the moduli K\"ahler potential admits a
limit in which the FI term has a value far below $M_{Pl}^2$, all
K\"ahler moduli can be stabilized even in the absence of
nonperturbative superpotential of $T$.

Since the generalization to the case with more moduli is straightforward,
here we consider a simple case with two K\"ahler moduli $T$ and $T^\prime$,
where $t=T+T^*$ corresponds to the volume of 4-cycle supporting the visible
sector, while $t^\prime=T^\prime+T^{\prime *}$ stands for hidden sector 4-cycle.
We assume that there exists a nonperturbative superpotential of $T^\prime$
generated by stringy instanton wrapping the hidden cycle, while no
nonperturbative term of $T$ due to the fermion zero modes on the visible sector
cycle.
In Type IIB string theory for KKLT compactification, the  K\"ahler potential of
$T_\alpha=\{T,T^\prime\}$ takes the no-scale form at the leading order
in the $\alpha^\prime$ and string loop expansion, so we consider a no-scale
K\"ahler potential obeying
\bea
K_0(\gamma t_A, \gamma t^\prime) =
K_0(t_A, t^\prime)-3\ln \gamma
\eea
for arbitrary real constant $\gamma$, where $t_A=t-\delta_{GS}V_A$.
We further assume that there exists a solution in the moduli space with
vanishing FI term:
\bea
\partial_t K_0 =0,
\eea
and explore the (local) minimum of the scalar potential near this
solution\footnote{
A simple example of such K\"ahler potential is the
one discussed in \cite{Stability wall}:
$
K_0=-\ln(
(t_1-\delta_{GS}V_A)(t_2+\delta_{GS}V_A)^2 +
\frac{1}{6}(t_2+\delta_{GS}V_A)^3 ),
$
where $t=t_1-t_2$, $t^\prime=t_1+t_2$, and the K\"ahler cone is defined
by $t_{1,2}>0$.
Note that $\xi_{FI}=\delta_{GS}(\partial_1K_0-\partial_2 K_0)/2=0$
on the wall in the K\"ahler cone defined by $t_2=4t_1$.
}.

In the absence of nonperturbative effects breaking the axionic shift
symmetry $T\rightarrow T+i\beta$ ($\beta=$ real constant), the $U(1)_A$
symmetry leads to an anomalous global symmetry
\bea
U(1)_{PQ}: \quad \phi_i\rightarrow e^{i\alpha} \phi_i,
\eea
which can be identified as a PQ symmetry solving the strong CP problem.
Then, this PQ symmetry should be broken spontaneously by the VEV of SM
singlet but $U(1)_A$ charged matter fields at a scale between $10^9$GeV
and $10^{12}$GeV.
As for those matter fields, we consider the example of the previous subsection.
Then the K\"ahler potential and superpotential of the model are given by
\bea
K &=& K_0(t_A,t^\prime) + X^*e^{-2V_A}X+ Y^*e^{6V_A}Y,
\nonumber \\
W &=& \omega_0 +  Ae^{-aT^\prime} + \lambda\frac{X^3Y}{M_{Pl}}
+ y_\Phi X\Phi\Phi^c.
\eea
Here we assume for simplicity that the matter K\"ahler metric of $X$ and $Y$
are independent of moduli, however our results equally apply to the case with
moduli-dependent matter K\"ahler metric.
For a no-scale moduli K\"ahler potential, we have
\bea
K^{\alpha\bar\beta}_0\partial_{\bar \beta}K_0 =
-(T^\alpha+T^{\alpha*}), \quad K^{\alpha\bar\beta}_0
(\partial_\alpha K_0)\partial_{\bar\beta}K_0 = 3,
\eea
and
\bea
2(\partial_t K_0)\partial_t K^{T\bar T^\prime}_0 +
(\partial_{t^\prime} K_0)\partial_t K^{T^\prime \bar T^\prime}_0
=0,
\eea
where $K^{\alpha\bar \beta}_0$ is the inverse of the moduli K\"ahler metric
$K_{0\alpha\bar\beta}=\partial_\alpha\partial_{\bar \beta}K_0$.
With these relations, one can find
\bea
\partial_t \left(V_{\rm SUGRA}+V_{\rm lift}\right) &=&
\left( V_F + \frac{2}{3}V_{\rm lift}
- 2\frac{\partial_t K^{TT^\prime}_0}{\partial_{t^\prime}K_0}
e^K|W_{T^\prime}|^2 \right)\partial_t K_0
\nonumber \\
&& +\, \frac{\delta_{GS}}{2}(\partial^2_t K_0)g^2_AD_A + (\partial_t
\ln g^2_A)\frac{g^2_A}{2}D^2_A,
\eea
where
\bea
D_A = \frac{\delta_{GS}}{2}\partial_t K_0+|X|^2-3|Y|^2.
\eea
The stationary condition $\partial_t(V_{\rm SUGRA}+V_{\rm lift})=0$ is thus
satisfied by the field configuration satisfying
\bea
\partial_t K_0 = 0, \quad |X|^2=3|Y|^2,
\eea
for which $D_A=0$.
On the other hand, as in the original KKLT scenario, the stationary condition
$\partial_{t^\prime}(V_{\rm SUGRA}+V_{\rm lift})=0$ leads to
\bea
D_{T^\prime}W = \frac{(\partial_{t^\prime}K_0)W}{\ln(M_{Pl}/m_{3/2})}
\left( 1 + {\cal O}\left(\frac{1}{\ln(M_{Pl}/m_{3/2})}\right) \right).
\eea
We then find $t$ and $t^\prime$ are fixed by the conditions
\bea
\partial_t K_0=0, \quad
\partial_{t^\prime}K_0 = \frac{aAe^{-aT^\prime}}{w_0}
\left( 1 + {\cal O}\left(\frac{1}{\ln(M_{Pl}/m_{3/2})}\right)\right),
\eea
and $X$ and $Y$ are fixed at
\bea
|X|^2=3|Y|^2\simeq
\frac{\sqrt3}{9|\lambda|}m_{3/2}M_{Pl},
\quad {\rm Arg}(X^3Y)={\rm Arg}(\lambda^*\omega_0)+\pi.
\eea
At this minimum, the SUSY breaking auxiliary components have VEVs as
\bea
&& \frac{F^X}{X}\,=\,\frac{F^Y}{Y}
\,\simeq\, -\frac{2}{3}m_{3/2}, \quad D_A = 0,
\nonumber \\
&& \frac{F^T}{T+T^*} \,\simeq\,
\frac{F^{T^\prime}}{T^\prime+T^{\prime*}}
\,\simeq\, \frac{m_{3/2}}{\ln(M_{Pl}/m_{3/2})},
\eea
where the relation between $F^T$ and $F^{T^\prime}$ is derived from
the no-scale relation
$K^{\alpha \bar \beta}_0 \partial_{\bar \beta}K_0=-(T^\alpha+T^{\alpha*})$.
Therefore, this variant of KKLT setup gives rise to a mixed modulus-gauge-anomaly
mediation\footnote{
It is worth noting that, if it is not charged under $U(1)_A$, $T$ can
alternatively be stabilized by the uplifting potential
\cite{Stabilization-by-uplifting}.
This gives a similar pattern of SUSY breaking, while the axion scale is
around the GUT or string scale.
}:
\bea
m_{\rm soft}({\rm MM}) \sim m_{\rm soft}({\rm GM}) \sim
m_{\rm soft}({\rm AM}) \gg m_{\rm soft}({\rm D}),
\eea
which was dubbed as deflected (or axionic) mirage mediation
\cite{Axionic-mirage,Deflected-mirage}.
As in the model discussed in the previous subsection, the Goldstone boson from
spontaneously broken $U(1)_{PQ}$ can play the role of the QCD axion with the axion
scale given by $v_{PQ}\sim \sqrt{m_{3/2}M_{Pl}}$.

\section{Conclusion}

There can be various  sources of SUSY breaking in models with anomalous $U(1)$
gauge symmetry: the $U(1)$ $D$-term, the $F$-components of the Green-Schwarz
modulus $T$ and the chiral matter $X$ introduced to cancel the Fayet-Iliopoulos
term, and also the SUGRA auxiliary component of the order of $m_{3/2}$.
Then the visible sector soft masses generically receive a modulus-mediated
contribution of the order of $F^T$ and a $D$-term contribution of the order
of $\sqrt{D_A}$ as well as the anomaly-mediated contribution of the order of
$\frac{g^2}{8\pi^2}m_{3/2}$.
Most of the known (semi) realistic string models with $U(1)_A$ include also
exotic SM gauge charged matter fields $\Phi$, $\Phi^c$ which become massive
through the Yukawa coupling to $X$, and therefore play the role of messenger
for the gauge mediation of SUSY breaking by $F^X$.
In such case, soft masses also receive a gauge-mediated contribution
of the order of $\frac{g^2}{8\pi^2}\frac{F^X}{X}$.

In this paper,  we have examined the relative strength of these modulus,
gauge, anomaly and $D$-term mediations in a simple class of models,
and find that various different mixed mediation scenarios can be realized
depending upon the characteristics of the $D$-flat directions and how those
$D$-flat directions are stabilized.
A key quantity which would determine the characteristics of the $D$-flat
directions is the ratio between the St\"uckelberg contribution to
the $U(1)_A$ gauge boson mass-square and the Fayet-Iliopoulos term.
Our results suggest that although its accurate structure is quite
model-dependent, it is quite common that soft terms in models with
$U(1)_A$ are not dominated by a single mediation, but determined by
a proper mixture of the moduli, gauge, anomaly and $D$-term mediations.

\section*{Acknowledgement}

We thank A. Uranga for useful discussions.  KC is supported by the
KRF Grants funded by the Korean Government (KRF-2008-314-C00064 and
KRF-2007-341-C00010) and the KOSEF Grant funded by the Korean
Government (No. 2009-0080844).
 KSJ is supported by the JSPS Grant-in-Aid 21-09224 for JSPS
Fellows. KO is supported by the Grant-in-aid for Scientific Research
No.21740155 and No.18071001 from MEXT of Japan.

\end{document}